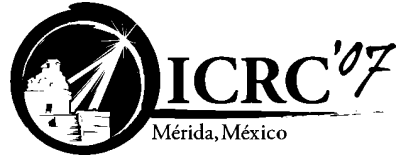

# Tunka-133 EAS Cherenkov Array: Status of 2007

N.M. BUDNEV[2], O.B. CHVALAEV[2], O.A. GRESS[2], N.N. KALMYKOV[1], V.A. KOZHIN[1],
E.E. KOROSTELEVA[1], L.A. KUZMICHEV[1], B.K. LUBSANDORZHIEV[3], R.R. MIRGAZOV[2], G. NAVARRA[5],
M.I. PANASYUK[1], L.V. PANKOV[2], V.V. PROSIN[1], V.S. PTUSKIN[4], Y.A. SEMENEY[2], A.V. SKURIKHIN[1],
B.A. SHAIBONOV (JUNIOR)[3], CH. SPIERING[6], R. WISCHNEWSKI[6], I.V. YASHIN[1], A.V. ZABLOTSKY[1],
A.V. ZAGORODNIKOV[2]

[1] *Scobeltsyn Institute of Nuclear Physics of Moscow State University, Moscow, Russia*
[2] *Institute of Applied Physics of Irkutsk State University, Irkutsk, Russia*
[3] *Institute for Nuclear Research of Russian Academy of Science, Moscow, Russia*
[4] *Institute of Terrestrial Magnetism, Ionosphere and Radiowave Propagation of Russian Academy of Science (IZMIRAN), Troitsk, Moscow Region, Russia*
[5] *Dipartimento di Fisica Generale Universita di Torino and INFN, Torino, Italy*
[6] *Deutsches Elektronen Synchrotron (DESY), Zeuthen, Germany*
kuz@dec1.sinp.msu.ru.

**Abstract:** The new EAS Cherenkov array Tunka-133, with about 1 km$^2$ sensitive area, is being installed in the Tunka Valley since the end of 2005. This array will permit a detailed study of the cosmic ray energy spectrum and the mass composition in the energy range of $10^{15}$ - $10^{18}$ eV with a unique method. The array will consist of 19 clusters, each composed of 7 optical detectors. The first cluster started operation in October 2006. We describe the data acquisition system and present preliminary results from data taken with the first cluster.

## Introduction

A thorough study of primary cosmic rays in the energy range of $10^{16}$ – $10^{18}$ eV is of utmost importance for understanding of the origin and propagation of cosmic rays in the Galaxy. It seems the maximum energy of cosmic rays accelerated in SN remnants to be in this energy domain [1]. As it was pointed out in [2], in this energy range the transition from Galactic to extragalactic cosmic rays may occur.

The new EAS Cherenkov array being under construction in the Tunka Valley (50 km from Lake Baikal), with 1 km$^2$ area, was named Tunka-133 [3,4]. It will allow to study cosmic rays by covering with a single method uniformly the energy range of $10^{15}$ - $10^{18}$ eV. During one year of operation (400 hours) Tunka-133 will record ~5·10$^5$ events with energy above 3·10$^{15}$ eV, ~300 events with energy higher than $10^{17}$ eV and a handful events with energy higher than $10^{18}$ eV.

Tunka Valley is famous for its good weather conditions (in winter time in particular). A number of EAS Cherenkov arrays – from Tunka-4 [5] to Tunka-25 [6,7] – have been operating at the site.

The new array will be methodically complementary to the "dense" 1 km$^2$ arrays KASCADE-Grande [8] and IceTop [9].

## The Tunka-133 array

The Tunka-133 array will consist of 133 optical detectors based on 8" EMI 9350 PMTs. 133 detectors are grouped in 19 clusters, each composed of seven detectors – six hexagonally arranged detectors and one in the center of the cluster. The distance between the detectors is 85 m. In addition to the Cherenkov detectors, 5-6 Auger-like (S = 10 m$^2$, depth = 90 cm) water tanks will be constructed for joint operation with the Cherenkov array. Two water tanks have been constructed up to now. The optimal distance between the water tanks is still under discussion.

The optical detector (fig.1) consists of a metallic cylinder with 50 cm diameter, containing a PMT. The container has a plexiglass window heated against frost at the top. The angular aperture is defined by the PMT shadowing. The efficiency is close to 100% up to 30° and reduces to 50% at zenith angles > 45°. The detector is equipped with remote controlled lid protecting the PMT from sunlight and precipitation. Besides the PMT with its high voltage supply and the preamplifiers, the detector box contains a light emitting diode for both amplitude and time calibration and a controller. The controller is connected with the cluster electronics via twisted pair by an RS-485 protocol. To provide the necessary dynamic range of $10^4$ two analog signals one from anode and another from a dynode are read out. They are amplified and then transmitted to the central electronics hut of each cluster. The ratio of amplitudes of these signals is about 30. It is not planned to heat the inner volume of the optical detector boxes, therefore all the detector electronics is designed to operate in a rather wide temperature range.

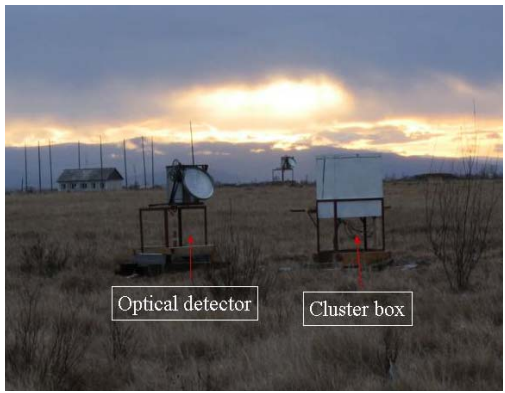

Figure 1: View on optical and cluster detector

The cluster electronics (fig.2) consists of the cluster controller, 4 blocks of four-channel FADCs, an adapter module connecting with optical detector controller and temperature controller. All electronic modules (except the temperature controller) are in VME standard. Each cluster is connected to the DAQ center through a multi-wire cable containing four copper wires and four optical fibers.

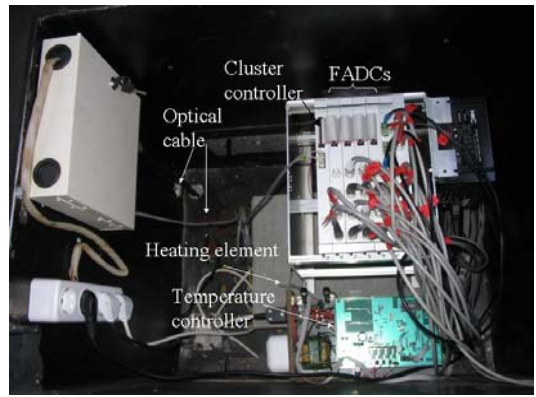

Figure 2: Inside view of cluster box.

The cluster controller contains optical transceiver, synchronization module, local time clock and cluster trigger module. Optical transceivers operating at 1000 MHz are responsible for data transmission and forming synchronization signals with a frequency 100 MHz for the cluster clock. The cluster trigger (the local trigger) is formed by a coincidence of at least three pulses from optical detectors above threshold within a time window of 0.5 μs. The time mark of the local trigger is measured by using the cluster clock. The accuracy of the time synchronization between different clusters is about 10 ns. The FADC modules are based on 12-bit 200 MHz ADCs and XILINX microchip FPGAs.

The temperature controller is used to maintain with the help of heating elements the necessary temperature inside of the container. Only when the temperature in the container becomes greater than 15 $^0$C, the temperature controller switches on power on VME crate.

## Preliminary data analysis

Algorithms for detectors calibration and EAS parameters reconstruction developed for Tunka-25 experiment have been adapted to the new array's geometry. The methods optimized for Tunka-133 data processing will finally provide an accuracy of about 6 m for the core location of ~15% for the energy measurements. We plan to incur the depth of the EAS maximum from the lateral distribution sharpness and the light pulse's width (FWHM) with accuracy better than ~25 g /cm$^2$.

Of course, we didn't reach this accuracy with data recorded with the first cluster of 7 detectors. The preliminary data allow us basically to check the cluster performance and to estimate threshold and event rate of the array.

The first cluster has operated for about 75 hours from November 2006 to January 2007. The trigger condition was the coincidence of at least three detector pulses above 100 p.e. The trigger rate is close to 0.2 Hz.

The apparatus has demonstrated stable operation during the whole winter. 16000 events overall have been recorded during 19 clear moonless nights. 5400 events have been reconstructed inside the cluster area (a circle a radius of 85 m) and within the solid angle limited by a zenith angle of 45°. The detectors signals are digitized with 5 ns step and 5000 ns depth.

After transmission via the preamplifier and cable RG-58 of 100 m length pulse has a front duration of about 15 ns and a FWHM of about 18 ns. The data processing procedure selects the points from the start of the front edge of the pulse to the end of the pulse tail. Usually we have more than 7 points for any pulse.

For large core distances (>300 m), the FWHM of EAS light pulse can vary from 20 ns to more than 100 ns. An example for waveforms from a very distant experimental event is presented in fig.3. The core distance and energy of this shower from the cluster center are estimated to be 600±100 m and (5±2)·10$^{17}$ eV correspondingly.

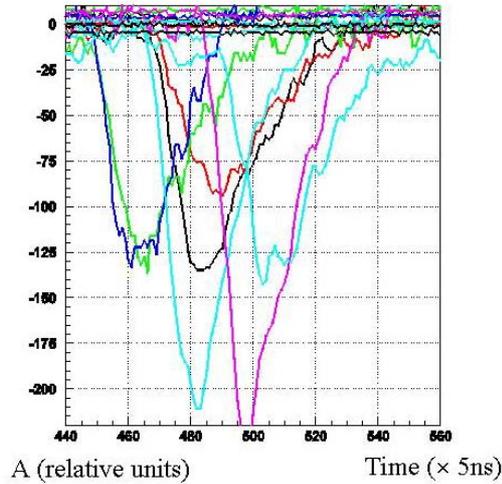

Figure 3: Example of a distant event. Pulses in different detectors are shown in different colors.

We estimate the arrival time of the Cherenkov light by extrapolating the pulse front to the constant fraction of 25% of the maximum pulse amplitude, and the total flux of light in the pulse by integration over the recorded wave form of the pulse.

An absolute time synchronization of 5 ns was achieved by equal lengths of all connecting cables. This accuracy was enough for the present preliminary data analysis. It will be improved later to an accuracy of about 1 ns.

The relative amplitude calibration was done by a method of comparing the density spectra recorded by different detectors. Using of this method in previous Tunka experiments was described in [10]. The absolute flux calibration was done by normalizing the obtained integral energy spectrum to the reference spectrum recorded with the QUEST experiment [11].

The integral energy spectrum of the recorded events is shown in fig. 4. A 100%-efficiency of the array is reached at an energy of about 10$^{15}$ eV. About 350 events have energies above 3·10$^{15}$ eV. This energy is used for comparison with the reference spectrum from the QUEST experiment [11]. The reference point is shown in fig. 4 by the black square. The spectrum is fitted with the different power laws for the energies below and above $3 \cdot 10^{15}$ eV.

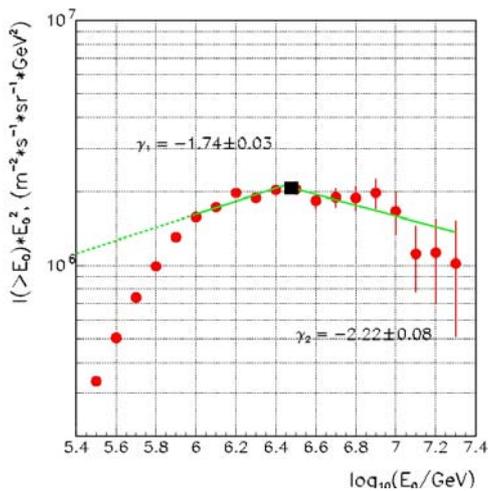

Figure 4: Preliminary integral energy spectrum taken with the first cluster of Tunka-133.

## Schedule of the array deployment

- Summer-Autumn 2006 – construction of the building for the central DAQ–system. Deployment of the first cluster (7 optical detectors). Operation of the first cluster during the winter season 2006 – 2007.
- Summer-Autumn 2007 – deployment of all cluster boxes and optical cables for all clusters. Deployment of 21 optical detectors. Operation of a four-cluster array (Tunka-28) during the winter season 2007-2008.
- Summer-Autumn 2008 – deployment of the main part of optical detectors and the construction of the third water tank. Commissioning of the array and start of operation.
- Summer-Autumn 2009 – deployment of the remaining part of optical detectors and construction of three water tank. Start of operation of the full array.

## Conclusion

A 1-km$^2$ Cherenkov EAS array is under construction in the Siberian Tunka Valley. The first cluster of the array has been successfully exploited during the last winter, proving reliability robustness of the array design Commissioning of the main part of the array is planned in autumn 2008.

## Acknowledgements

The present work is supported by the Russian Ministry of Education and Science, by the Russian Fund of Basic Research (grants 05-02-04010, 05-02-16136, 06-02-16526) and by the Deutsche Forschungsgemeinschaft DFG (436 RUS 113/827/0-1)